\documentclass[preprint2]{aastex}

\usepackage{color}
\definecolor{Dred}{rgb}{0.312,0.070,0.070}
\definecolor{Dblue}{rgb}{0.070,0.070,0.312}
\definecolor{Dgreen}{rgb}{0.070,0.312,0.070}
\definecolor{Db}{rgb}    {0.050,0.0,0.320}

\newcounter{note}
\let\oldmarginpar\marginpar
\renewcommand\marginpar[1]{\-\oldmarginpar[\raggedleft\footnotesize #1]{\raggedright\footnotesize #1}}

\shorttitle{First Detection of 350 Micron Polarization}
\shortauthors{Lee et al.}

\begin{document}

\title{First Detection of 350 Micron Polarization From a Radio-loud AGN}

\author{
Sang-Sung Lee\altaffilmark{1,2},      
Sincheol Kang\altaffilmark{1,2},      
Do-Young Byun\altaffilmark{1},        
Nicholas Chapman\altaffilmark{3},     
Giles Novak\altaffilmark{3},          
Sascha Trippe\altaffilmark{4},        
Juan Carlos Algaba\altaffilmark{1},   
and
Motoki Kino\altaffilmark{1}           
}

\altaffiltext{1}{Korea Astronomy and Space Science Institute, 
Daedeokdae-ro 776, Yuseong-gu, Daejeon 305-348, Korea; sslee@kasi.re.kr}
\altaffiltext{2}{Korea University of Science and Technology, 176 Gajeong-dong, Yuseong-gu, Daejeon 305-350, Korea}
\altaffiltext{3}{Center for Interdisciplinary Exploration and Research
in Astrophysics (CIERA) and Department of Physics \& Astronomy,
Northwestern University, 2145 Sheridan Road, Evanston, IL 60208, USA }
\altaffiltext{4}{Department of Physics and Astronomy, Seoul National University,
599 Gwanak-ro, Gwanak-gu, Seoul 151-742, Republic of Korea}

\begin{abstract}
We report the first detection of linearly polarized emission at
an observing wavelength of 350 $\mu$m
from the radio-loud Active Galactic Nucleus 3C 279.
We conducted polarization observations for 3C 279
using the SHARP polarimeter in
the Caltech Submillimeter Observatory on 2014 March 13 and 14.
For the first time,
we detected the linear polarization with the degree of polarization of
13.3\%$\pm$3.4\% ($3.9\sigma$) and the Electric Vector Position Angle (EVPA)
of 34.7$^\circ\pm5.6^\circ$.
We also observed 3C~279 simultaneously at 13, 7, and 3.5 mm
in dual polarization
with the Korean very long baseline interferometry (VLBI) Network on 2014 March 6 (single dish)
and imaged in milliarcsecond (mas) scales at 13, 7, 3.5, and 2.3 mm on March 22 (VLBI).
We found that the degree of linear polarization 
increases from 10\% to 13\% at 13 mm to 350 $\mu$m
and the EVPAs at all observing frequencies
are parallel within $<10^\circ$ to the direction of the jet at mas scale,
implying that the integrated magnetic fields are perpendicular
to the jet in the innermost regions.
We also found that the Faraday rotation measures RM are
in a range of $-6.5\times10^2 \sim -2.7\times10^3$ rad m$^{-2}$
between 13 and 3.5 mm, and are scaled as a function of wavelength: $|{\rm RM}|\propto\lambda^{-2.2}$. 
These results indicate that the mm and sub-mm polarization emission are generated
in the compact jet within 1~mas scale and affected by a Faraday screen
in or in the close proximity of the jet.
\end{abstract}


\section{Introduction}

Radio-loud active galactic nuclei (AGNs) are known to generate
highly collimated relativistic jets.
Polarimetric observations at millimeter wavelengths of the radio-loud AGNs find 
polarized synchrotron emission
with the degree of linear polarization 
in the range of 1\%--19\% with a mean value of 4\%~\citep{tri+10,agu+10}.
Polarization observations enable us to understand 
geometrical structure and intensity of magnetic fields,
particle densities and structures of
emission region~\citep[e.g.,][and references therein]{ss88}.

Synchrotron emission at millimeter (mm) and sub-millimeter (sub-mm) wavelengths
comes from the innermost compact regions of the relativistic jet,
most likely from within
the most upstream emission region in the compact radio jet (the core)
in Very Long Baseline Interferometry (VLBI)
observations~\citep[e.g.,][]{lee+08,jor+07}.
The mm and sub-mm emission is typically optically thin
and less affected by Faraday rotation
than at longer wavelengths~\citep[see e.g.,][]{agu+10}.
Since the amount of Faraday rotation (RM)
 is proportional to
the squares of observing wavelength ($\lambda$) as
$\chi _{\rm {obs}}=\chi _{\rm {int}}+\rm {RM} \lambda ^{2}$
(where  RM\footnote[1]{
RM in the observer's frame can be scaled by a factor of $(1+z)^2$
in the rest frame. In this letter, we present RM in the observer's frame. 
} is the Faraday rotation measure,
$\chi _{\rm obs}$ is the observed electric vector
linear polarization angle of a source,
and $\chi _{\rm int}$ is the intrinsic polarization angle),
mm and sub-mm polarization observations
may experience much less Faraday rotation
than centimeter wavelength observations.
Provided that the Faraday rotation measure is constant,
at $\lambda 350~\mu$m, a rotation measure of $1.4\times10^5$ rad ${\rm m^{-2}}$
causes a rotation of the Electric Vector Position Angle (EVPA) by $1^\circ$,
whereas it causes rotation by $3200^\circ$ at 2 cm.
Accordingly, such high rotation measures are studied best
at (sub)mm wavelengths where $\chi _{\rm {obs}}=\chi _{\rm {obs}}\pm n\pi$
ambiguities can be avoided.
This indicates that the mm and sub-mm observations
may be much more efficient to investigate the intrinsic
linear polarization properties of the relativistic jets
than centimeter radio observations.

Previous mm/sub-mm polarization observations include 
several polarimetric mm surveys of AGNs with the IRAM
Plateau de Bure Interferometer~\citep{tri+10} and the IRAM 30 m
telescope~\citep{agu+10,agu+14},
and polarimetric studies of individual AGNs~\citep{ste+96,jor+05,jor+07,tri+12a,tri+12b,pla+14,mart+15}.
These studies report that degree of linear polarization $p$ of
the radio loud AGNs is in the range of 1\%-19\% at mm/sub-mm wavelengths.

Among the radio loud AGNs in these studies,
3C 279 (1253-055, $z=0.538$) is one of the brightest, highly polarized AGNs. 
In 2010 August, as reported by \cite{agu+14}, 
the degree of linear polarizations of 3C~279 was $3.9\%\pm0.3\%$ at 3.5 mm
and of $<6.6\%$ at 1.3 mm. The EVPA was measured only at 3.5 mm to be
$107^\circ\pm1.8^\circ$.
\cite{ste+96} showed that 3C~279 was highly 
polarized with $p=10.4\%\pm0.6\%$ and $p=13.5\%\pm1.4\%$ at 1.1 and 0.8 mm.
The flux densities of 3C 279 were $7.6\pm0.5$ Jy and $5.3\pm0.4$ Jy
at each wavelengths on 1995 August 2.
The polarization position angle of $76^\circ$ at both 1.1 and 0.8 mm implies
that the magnetic field is aligned orthogonal to the centimeter VLBI jet
to within about $6^\circ$.
\cite{jor+07} also reported that 3C~279 was the most highly polarized AGNs
among their sources with 
$p=3\%-9\%$ at 1.3 mm in 1998-1998 and $p=7\%-11\%$ at 0.8 mm in 2000-2001,
and with the mean polarization position angle $67^\circ$ at 1.3 mm
and $65^\circ$ at 0.8 mm.
These results clearly show that mm/sub-mm polarization
of 3C~279 is significantly variable.

In order to investigate intrinsic linear polarization properties
of the relativistic jet of 3C~279, we conducted
(quasi-)simultaneous multiwavelength
polarization observations
at 1.3 cm, 7 mm, 3.5 mm, and 350 $\mu$m.
In this letter, we present results of mm and sub-mm
observations of 3C 279 using the Korean VLBI Network (KVN) and
the Caltech Submillimeter Observatory (CSO),
which yielded the first detection of linearly polarized emission
of 3C 279 at an observing wavelength of 350 $\mu$m.
Combining the KVN and CSO polarization observations, 
we aim to study the magnetic field properties
in the innermost regions of the jet
and their environments. 
Throughout this letter,
we assume a cosmology with $H_0 = 70~{\rm km}~{\rm s}^{-1}~{\rm Mpc}^{-1}$,
$\Omega_M = 0.3$, and $\Omega_{\lambda} = 0.7$.

\section{Observations and data reduction}\label{sec2}

We conducted polarimetric observations of 3C~279 on 2014 March 13 and 14 UT
using the SHARP polarimeter at the CSO.
SHARP is a fore-optics module that adds polarimetric
capability to SHARC-II, a $12\times32$ pixel bolometer array used
at the CSO~\citep{dow+03,li+08}.
The wavelength of the observations was $350~\mu$m.
We observed 3C~279 for 30 half-wave plate (HWP) cycles,
where one HWP cycle consists of the observations of the
source at four HWP angles and takes about 7 minutes.
The chop throw was $3'$ or $5'$ and
the optical depth at $350~\mu$m ranged from 0.90 to 2.6
(or 0.05--0.1 at 225~GHz).
Details about SHARP and data reduction procedures can be found
in \cite{dav+11}  and \cite{cha+13}.
Following these procedures, we created maps of Stokes $I$ (total
intensity) and Stokes $Q$ and $U$ (linear polarization).

We also conducted single dish
simultaneous multifrequency (22, 43, and 86~GHz)
polarimetric observations of 3C~279
using 21 m radio telescopes of
KVN on 2014 March 6, 
and conducted high resolution
simultaneuos multifrequency (22, 43, 86, and 129~GHz)
VLBI observations of the source on 2014 March 22.
KVN is a mm-dedicated VLBI network consisting of
three 21 m radio telescopes in Korea with the maximum
baseline of 476 km~\citep{lee+11,lee+14}.

A polarimetric observation of the source
consists of on--off switching observations
with the on-source integration time of $\sim$6 minutes.
Cross-scan pointing and antenna gain calibration measurements were conducted
before every polarimetric observation.
We observed 
planets for calibrating instrumental polarization,
and Crab nebula as a polarization angle calibrator.
The phase difference between left and right circularly polarized signals is
derived from the complex cross-power spectra
measured with
the KVN digital spectrometer.
The complex cross-power spectra are corrected for
instrumental polarization using those of planets.
The phase of the cross-power spectra are corrected for a phase rotation due to
parallactic angle change and then used to estimate
the polarization angle.
The estimated polarization angle is corrected
by $\chi=154^{\circ}$ at 22 GHz, 43 GHz, and 86 GHz bands,
which is the intrinsic $\chi$ of Crab nebula~\citep{wf80,fh79,aum+10}.
The rms uncertainties of the linear polarization observations
are about 10 mJy and 15 mJy at 22 GHz and 43 GHz, respectively.
The systematic error of the polarization angle measurements
is $2^\circ$ at both frequencies.
The typical instrumental polarization leakage of the KVN system
is $<5\%$ at both frequencies.
Details about the data reduction pipeline for the polarimetric observations
and the polarimetric capability of the KVN
will be described elsewhere (Byun, D.-Y. et al. 2015, in preparation).

The VLBI observations of the source were conducted
as part of the iMOGABA project~\citep{lee+13}.
The source was observed with 3 scans of 5 minutes long each
during UT 12:30:00 -- 15:05:00.
The observing frequency band was 22.700--22.764 GHz, 43.400--43.464 GHz,
86.800--86.864 GHz, 129.300--129.364 GHz
in left-hand circular polarization.
We conducted sky tipping curve measurements
every hour in order to trace changes
in the opacity of the atmosphere during the observations.
The data were digitized in 2-bit sampling mode,
and the digitized signals were
processed by the digital filter bank
to be 16 sub-bands of 16 MHz and divided evenly for four frequency bands.
The Mark 5B system at a recording rate of 1024 Mbps
was used for recording the data.
The correlation of the recorded data
was done with the DiFX software correlator.
We followed the standard data reduction procedure for the KVN observations
as described in \cite{lee+14}, and obtained the high resolution
VLBI images for the source.

The 1~mm (2014 March 7) and 850 $\mu$m (2014 March 20) flux density data
were obtained at the Submillimeter Array (SMA), an eight-element interferometer
located near the summit of Mauna Kea (Hawaii). 3C 279 is included in
an ongoing monitoring program at the SMA to determine the flux densities
of compact extragalactic radio sources that can be used as calibrators at mm and sub-mm wavelengths~\citep{gur+07}.
Observations of calibrators were conducted every 3-5 minutes,
and the measured source signal strength calibrated against
known standards, typically solar system objects
(Titan, Uranus, Neptune, or Callisto).
Data from this program are updated regularly and are available at the SMA
website (http://sma1.sma.hawaii.edu/callist/callist.html).

Optical (5000-7000~\AA) spectropolarimetric data was obtained
at the Steward Observatory, as part of a monitoring program
of gamma-ray blazars.
The data used in this paper is taken on 2014 March 25.
The detailed procedure of data analysis is described in \cite{smi+09}.

\section{Results}\label{sec3}

\begin{figure}[t!]
\epsscale{1.0}
\plotone{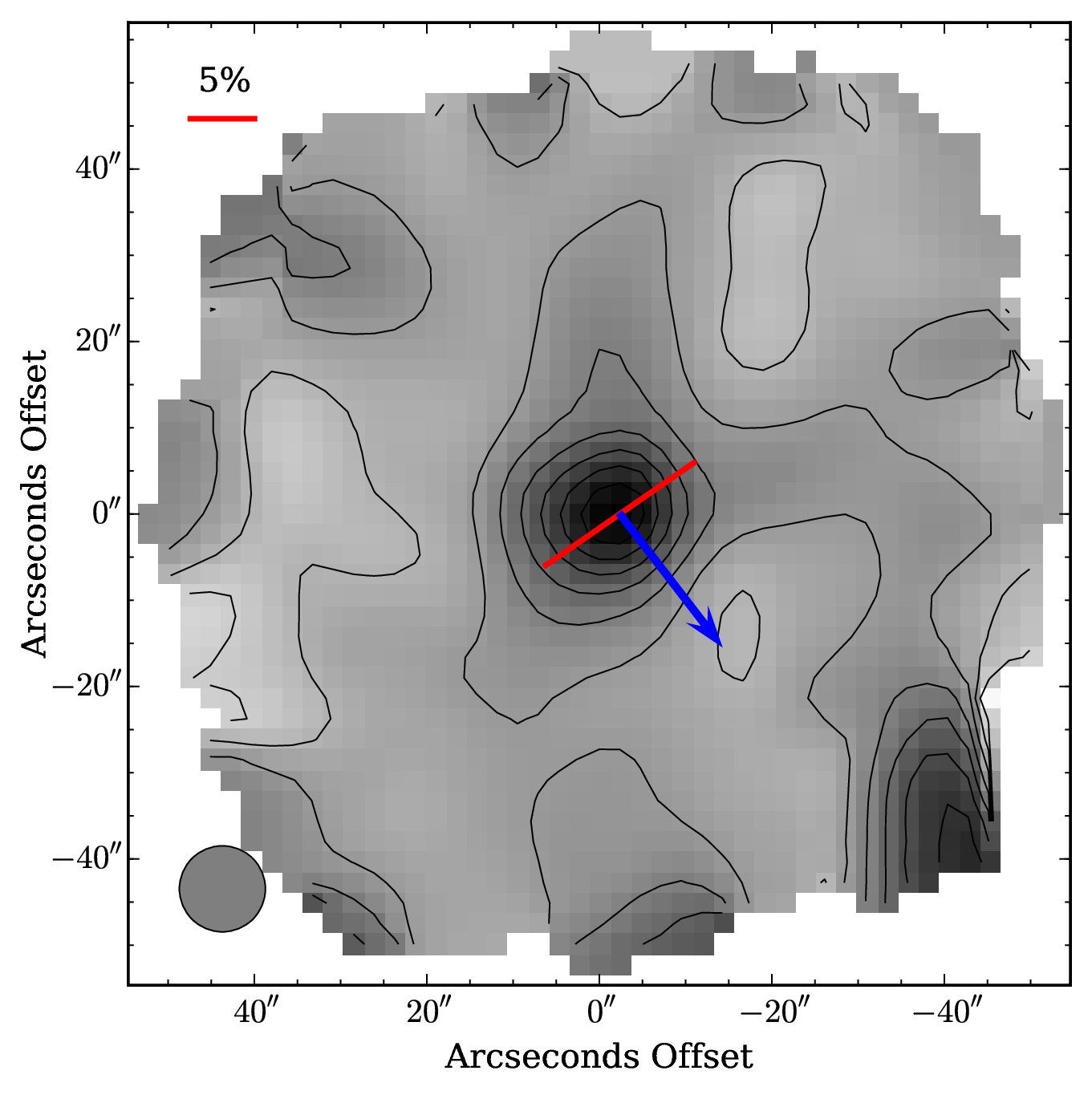}
\caption{Polarization map of 3C~279. Contours show the $350~\mu$m
intensity (Stokes I),
ranging from 20\% of the peak flux, in steps of 10\%. 
The red vector shows the measured $350~\mu$m polarization at the peak,
where the angle of the vector has been rotated by $90^\circ$ to show
the inferred magnetic field direction.
The length of the vector is proportional to
the degree of polarization.
The gray circle at the bottom denotes the SHARP beam size, $10^{''}$.
The large blue arrow shows the mean position angle (-142.2$^{\circ}$)
of the jets observed at 22-129 GHz using KVN.
\label{fig-3c279}}
\end{figure}

\begin{figure}[th!]
\epsscale{1.0}
\plotone{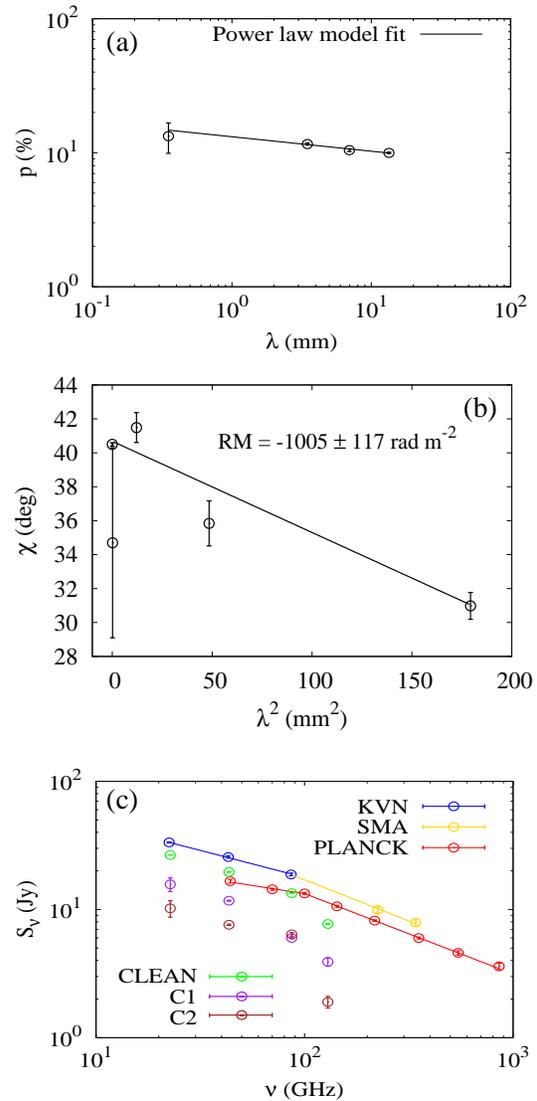}
\caption{
(a) Degree of linear polarization of 3C 279 at 13 mm, 7 mm, 3.5 mm, and 350 $\mu$m.
The solid line is the best fit result of the power law model to the data.
(b) EVPAs of 3C~279 at  13 mm, 7 mm, 3.5 mm, 350 $\mu$m, and optical.
The solid line is the best fit result of the Faraday rotation measure RM to the data.
(c) Flux densities of 3C~279 obtained with KVN single dish (blue circles), SMA (yellow circles), PLANCK (red circles),
and KVN VLBI observations (green circles for CLEAN images,
purple circles for C1 component and brown circles for C2a or C2b component). The solid lines are the best fit results of the spectral index to the data.
\label{fig-3c279all}}
\end{figure}

\begin{figure*}[th]
\epsscale{2}
\plotone{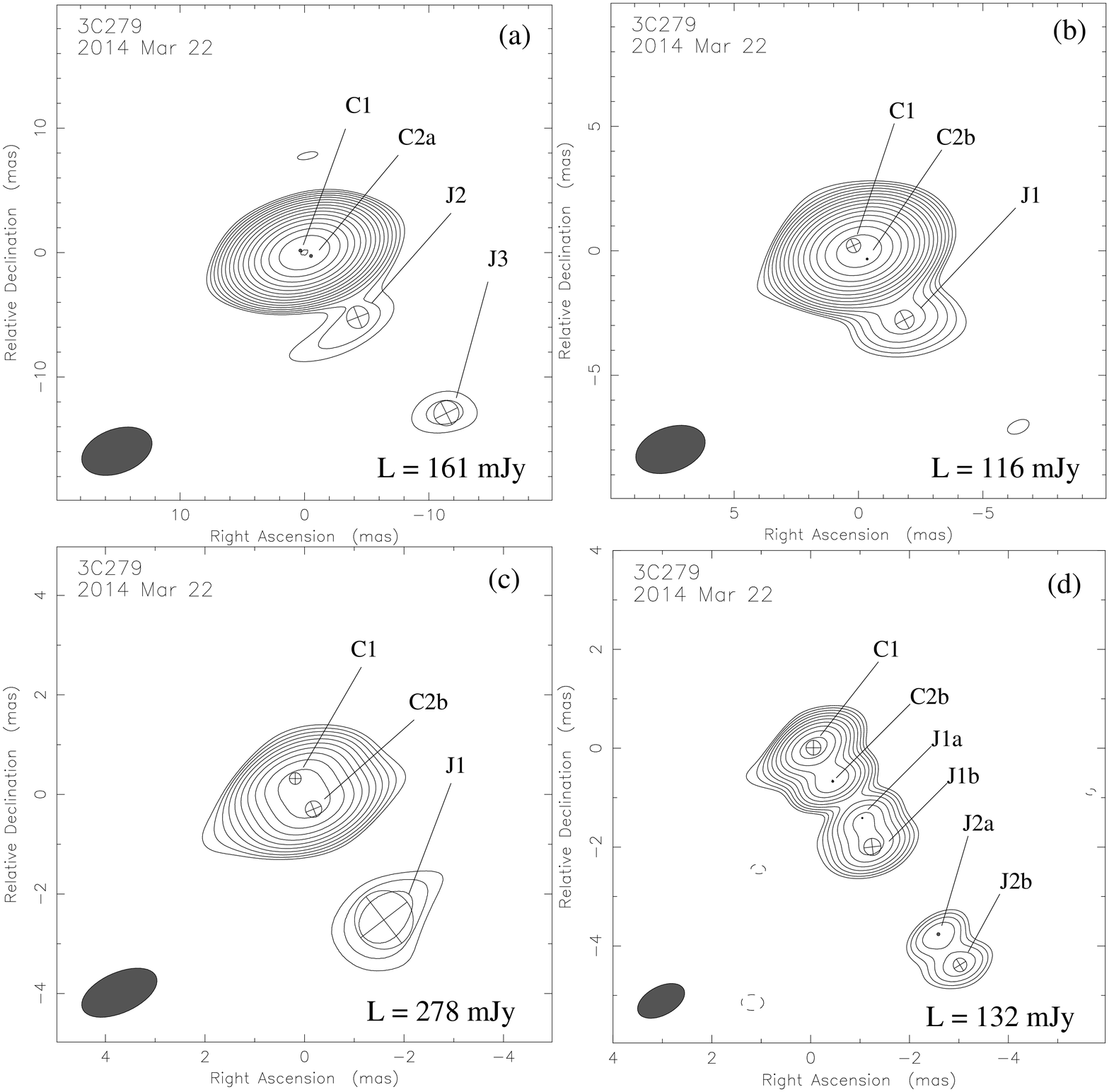}
\caption{
CLEANed images of 3C~279 obtained with KVN
        at (a) 13 mm, (b) 7 mm, (c) 3.5 mm, and (d) 2.3 mm,
	on 2014 March 22.
        Circular Gaussian models are on top of the contour map.
        The axes of each map are the relative R.A. and decl. offsets
        from the tracking center in mas.
        The lowest contour level is shown in the lower right corner of
        each map. The contours have a logarithmic spacing and
        are drawn at 1, 1.4, ..., $1.4^n$ of the lowest contour level.
\label{fig-kvnmap}}
\end{figure*}

\begin{deluxetable}{lccccccccc}
\tabletypesize{\scriptsize}
\tablecaption{Results of the polarization observations\label{t1}}
\tablewidth{0pt}
\tablehead{
\colhead{} & \colhead{} &\colhead{$\lambda$} & \colhead{$\nu$} & \colhead{$p$} & \colhead{$\sigma_{p}$} & \colhead{$\chi$} & 
\colhead{$\sigma_{\chi}$} & \colhead{$S_{\nu}$}\\
\colhead{Epoch} & 
\colhead{Telescope} & 
\colhead{(mm)} & 
\colhead{(GHz)} & 
\colhead{(\%)} &
\colhead{(\%)} & 
\colhead{(deg)} & 
\colhead{(deg)} & 
\colhead{(Jy)}\\
\colhead{(1)} & 
\colhead{(2)} & 
\colhead{(3)} &
\colhead{(4)} & 
\colhead{(5)} & 
\colhead{(6)} & 
\colhead{(7)} & 
\colhead{(8)} & 
\colhead{(9)}
}
\startdata
2014 Mar 6     & KVN  &13   & 22.4 & 10.0  & 0.1 & 31.0 & 0.8 & $33.4\pm0.3$\\
               &  &7.0  & 43.1 & 10.5 & 0.2 & 35.8 & 1.3 & $25.6\pm0.5$\\
               &  &3.5  & 86.2 & 11.6 & 0.2 & 41.5 & 0.9 & $18.8\pm0.5$\\
2014 Mar 13 and 14 & CSO &0.35 & 860  & 13.3  & 3.4  & 34.7  & 5.6  & ...\\
2014 Mar 25    & Steward & 5000-7000\AA   & ...     & 12.4 & 0.04  & 40.5 & 0.1  & ...\\
\enddata
\tablecomments{ 
Column designation:
(1)~-~observation epoch;  
(2)~-~telescope;
(3)~-~observing wavelength in mm;
(4)~-~observing frequency in GHz;
(5)~-~degree of linear polarization;
(6)~-~standard deviation of the degree of linear polarization;
(7)~-~Electric Vector Position Angle (EVPA) in degree;
(8)~-~standard deviation of the EVPA;
(9)~-~flux density in Jansky.
}
\end{deluxetable}

\begin{deluxetable}{lcccccccccrcr}
\tabletypesize{\scriptsize}
\setlength{\tabcolsep}{0.1cm}
\tablecaption{Image parameters\label{table-par}}
\tablewidth{0pt}
\tablehead{
\colhead{$\lambda$} &
\colhead{$B_{\rm maj}$:$B_{\rm min}$:$B_{\rm PA}$} &
\colhead{$S_{\rm CLEAN}$}  &
\colhead{$S_{\rm p}$} & 
\colhead{$\sigma$} &
\colhead{Comp} &
\colhead{$S_{\rm tot}$}  &
\colhead{$S_{\rm peak}$}  &
\colhead{$d$}  &
\colhead{$r$}  &
\colhead{$\theta$} &
\colhead{$r^{\prime}$}  &
\colhead{$\theta^{\prime}$}\\
\colhead{(1)} &
\colhead{(2)} &
\colhead{(3)} &
\colhead{(4)} & 
\colhead{(5)} &
\colhead{(6)} &
\colhead{(7)} &
\colhead{(8)} &
\colhead{(9)} &
\colhead{(10)} &
\colhead{(11)} &
\colhead{(12)} &
\colhead{(13)}
}
\startdata
13 mm  & 5.88:3.56:-70.2 & 26.6 & 25.4 & 53.7 & C1 & $15.7\pm1.9$   & $15.8\pm1.4$   & $0.23\pm0.03$ & $0.37\pm0.02$ & $63.3\pm2.5$  & 0.00 & 0.0 \\ 
       &                 &      &      &      & C2a& $10.2\pm1.5$   & $10.3\pm1.1$   & $0.23\pm0.05$ & $0.59\pm0.02$ & $-117.2\pm2.3$& 0.96 & -117.0 \\ 
       &                 &      &      &      & J2 & $0.295\pm0.149$& $0.288\pm0.104$& $1.80\pm0.65$ & $6.73\pm0.33$& $-140.5\pm2.8$& 7.07 & -139.3 \\ 
       &                 &      &      &      & J3 & $0.258\pm0.165$& $0.265\pm0.118$& $2.02\pm0.90$ & $17.23\pm0.45$& $-138.5\pm1.5$&17.57 & -138.1  \\ 
[0.1cm]
7 mm   & 2.90:1.79:-70.4 & 19.6 & 17.2 & 38.6 & C1 & $11.7\pm0.1$   & $11.0\pm0.1$   & $0.58\pm0.00$ & $0.29\pm0.00$ & $43.4\pm0.4$  & 0.00 &  0.0 \\ 
       &                 &      &      &      & C2b& $7.65\pm0.1$   & $7.64\pm0.07$  & $0.08\pm0.00$ & $0.48\pm0.00$ & $-133.4\pm0.0$& 0.77 & -134.6 \\ 
       &                 &      &      &      & J1 & $0.624\pm0.053$& $0.556\pm0.035$& $0.78\pm0.05$ & $3.33\pm0.02$ & $-146.3\pm0.4$& 3.62 & -145.5 \\ 
[0.1cm]
3.5 mm & 1.62:0.81:-66.1 & 13.4 & 7.57 & 92.7 & C1 & $6.00\pm0.12$ & $5.73\pm0.08$ & $0.24\pm0.00$ & $0.37\pm0.00$ & $30.1\pm0.3$  & 0.00 &  0.0 \\ 
       &                 &      &      &      & C2b& $6.36\pm0.16$ & $5.81\pm0.11$ & $0.33\pm0.01$ & $0.35\pm0.00$ & $-148.4\pm0.5$& 0.72 & -149.2 \\ 
       &                 &      &      &      & J1 & $1.52\pm0.68$ & $0.52\pm0.22$ & $1.18\pm0.50$ & $2.98\pm0.25$ & $-147.7\pm4.7$& 3.35 & -147.9 \\ 
[0.1cm]
2.3 mm & 1.03:0.58:-52.2 & 7.74 & 3.35 & 44.0 & C1 & $3.91\pm0.29$ & $3.35\pm0.19$ & $0.30\pm0.02$ & $0.06\pm0.01$ & $-79.4\pm8.3$ & 0.00 &  0.0 \\ 
       &                 &      &      &      & C2b& $1.87\pm0.16$ & $1.87\pm0.12$ & $0.04\pm0.00$ & $0.81\pm0.00$ & $-146.3\pm0.1$& 0.79 & -150.3 \\ 
       &                 &      &      &      & J1a& $0.988\pm0.10$& $0.99\pm0.07$ & $0.03\pm0.00$ & $1.76\pm0.00$ & $-143.5\pm0.1$& 1.73 & -145.3 \\ 
       &                 &      &      &      & J1b& $1.34\pm0.20$ & $1.11\pm0.13$ & $0.35\pm0.04$ & $2.35\pm0.02$ & $-148.0\pm0.5$& 2.33 & -149.4 \\ 
       &                 &      &      &      & J2a& $0.483\pm0.11$& $0.49\pm0.08$ & $0.06\pm0.02$ & $4.56\pm0.01$ & $-145.5\pm0.1$& 4.54 & -146.2 \\ 
       &                 &      &      &      & J2b& $0.544\pm0.06$& $0.48\pm0.04$ & $0.27\pm0.02$ & $5.32\pm0.01$ & $-145.4\pm0.1$& 5.30 & -146.0 \\ 
\enddata
\tablecomments{ 
Column designation:
(1)~-~Observing wavelength;
(2)~-~restoring beam-major axis (mas): minor axis (mas): position angle of the major axis ($^\circ$);
(3)~-~total CLEAN flux density (Jy); 
(4)~-~peak flux density in the image (Jy~beam$^{-1}$); 
(5)~-~off-source RMS in the image (mJy~beam$^{-1}$);
(6)~-~Gaussian jet components;
(7)~-~model flux density of the component (Jy); 
(8)~-~peak brightness of individual component measured in the image (mJy~beam$^{-1}$); 
(9)~-~size (mas); 
(10)~-~radius (mas); 
(11)~-~position angle ($^\circ$);
(12)~-~radius with respect to the C1 component (mas); 
(13)~-~position angle with respect to the C1 component ($^\circ$).
}
\end{deluxetable}

Our final map at $350~\mu$m (Figure~\ref{fig-3c279}) obtained using the SHARP
has a resolution of $10''$.
Because 3C~279 is a point source at our resolution,
we only extracted a polarization vector at the position of
the peak flux. Degree of polarization cannot be negative,
which leads to a small positive bias,
for which we corrected~\citep{hil+00}.
After debiasing, the degree of linear
polarization $p$ is $13.3\% \pm 3.4\%$ ($3.9\sigma$), and
the polarization angle $\chi$ is $34.7^\circ \pm 5.6^\circ$ measured
east of north. This is the first detection of linearly polarized emission
at the observing wavelength of $350~\mu$m from AGNs.
The polarization observations using the KVN telescopes
at mm wavelengths, 13 mm, 7 mm, and 3.5 mm,
have been processed with the KVN pipeline,
and resulted in the degree of linear polarization $p=10\%$--$12\%$
and the polarization angle $\chi=32^\circ$--$41^\circ$.
From the optical spectropolarimetric observations,
we found that the optical polarization degree is $12.35\%\pm0.05\%$
and the polarization angle is $\chi=40.5^\circ\pm0.1^\circ$.
We summarized the results of the polarization observations in Table~\ref{t1}.

From the mm and sub-mm polarization observations,
we found that the degree of linear polarization 
increases from 10\% to 13\% as the observing wavelength becomes shorter
and the polarization angle rotates
from $30^{\circ}$ to $41^{\circ}$ to east from north
and at $350~\mu$m the angle rotates back to $35^{\circ}$,
as shown in Figures~\ref{fig-3c279all}a and \ref{fig-3c279all}b.
With the polarization angles observed at multiwavelengths,
we estimated RM of $-647\pm206$ rad m$^{-2}$ between 7-13~mm,
$-2713\pm766$ rad m$^{-2}$ between 3.5-7~mm
and $-1022\pm264$ rad m$^{-2}$ between 3.5 and 13 mm.  
By assuming that the polarized emission at radio to optical
are coming from a very compact region (or the same region),
we may be able to estimate the RM of $-1005\pm117$ rad m$^{-2}$.

The flux densities at 13-3.5~mm were $>18$ Jy with a spectral index
of $\alpha=-0.42\pm0.01$ ($S_{\nu}\propto\nu^{\alpha}$).
The flux densities obtained with SMA are
$9.9\pm0.5$ Jy at 1~mm on 2014 March 7,
and $7.9\pm0.4$ Jy at 850 $\mu$m on 2014 March 20,
yielding a spectral index of $\alpha=-0.64\pm0.02$.
This implies that 3C~279 is a flat spectrum source
at cm and mm wavelengths, and its spectrum becomes steeper
at mm and sub-mm wavelengths as shown in Figure~\ref{fig-3c279all}c.
We obtained the {\it Planck} flux density of 3C~279
from the recently released \cite{pla13} catalog, which yielded
comparable spectral indices of $-0.25\pm0.04$ and $-0.63\pm0.01$
at the frequency ranges of 44-100 and 100-857~GHz, respectively,
with those obtained from KVN and SMA observations.
We found that the spectrum of the source becomes steeper at $>$100~GHz.
The difference of the flux densities at the similar frequencies
between the KVN/SMA and the PLANCK spectra is mainly due to 
the source variability.

In Figure~\ref{fig-kvnmap}, we present
CLEANed images for 3C~279 obtained with KVN observations simultaneously
at 13, 7, 3.5, and 2.3 mm on 2014 March 22.
Circular Gaussian models are fitted to the CLEANed images.
The VLBI images at mm wavelengths show 
core-dominated structures within a few mas and faint jet components
in south-west direction.
More detailed parameters of the images are summarized in Table~\ref{table-par}.
The uncertainties of the parameters are estimated
by following~\cite{lee+08}.
We found that two comparable jet components separated by $<$1 mas, i.e.,
the C1 and C2a(b) components, and they are the brightest jet components
in the images at all wavelengths.
All jet components are aligned in a mean position angle of $-142.4^{\circ}$, 
taking into account the estimated jet position angle with respect to
the C1 component.
The direction of the jets is aligned to the polarization position angle
at mm-to-optical within $<10^{\circ}$.
This implies that the dominant magnetic field direction is perpendicular
to the direction of the mm jet at mas scales.

\section{Discussion and Conclusions}\label{sec4}

High resolution VLBI observations of 3C~279
show that the combined flux density of the C1 and C2a(b) components
contributes over 90\% of the total CLEAN flux density at 13-3.5 mm,
and over 70\% of that at 2.3 mm,
implying that
most of mm emission originate from very compact regions
within $<$1 mas ($<$6.3 pc).
Previous VLBI observations also confirmed 
the compact jet structures~\citep[e.g.,][]{jor+05,lee+08,hom+09}.
In addition to the compactness of the source,
our KVN VLBI observations at multifrequency revealed the spectral properties of
the innermost regions, i.e., the components C1 and C2.
As shown in Figure~\ref{fig-3c279all}c, there is a break at 86 GHz on
the CLEAN spectrum of the source, implying that the source becomes
optically thin above the frequency on mas scales.
The spectral break seems mainly due to the spectrum of the component C2
which shows very steep spectrum between 86 and 129 GHz.
Therefore, we may expect that the component C1 dominates
the sub-mm emission and hence the sub-mm ($350~\mu$m) emission regions
are located within the C1 component.

In order to further investigate the mm and sub-mm polarized emission
predominantly originated in the components C1 and C2 (i.e., within $<$1~mas), 
the degree of linear polarization at 13 mm-$350~\mu$m
was fitted with a power law model, $p(\%)=A\lambda^{\beta}$, suggested by
\cite{tri91} and modified by \cite{far+14}
for explaining external Faraday depolarization~\citep[see][]{burn66}, 
where $A$ is constant, $\beta$ is a polarization spectral index,
and $\lambda$ is an observing wavelength in cm.
A best fit power law to the data yielded $A=10.3\pm0.08$ and
$\beta=-0.11\pm0.01$
which implies that the effect of external Faraday depolarization
is very small within the observing beams of KVN and CSO.
We also expect that our measurements of the polarization
with different angular resolutions ($10''$-$130''$)
are not significantly affected by beam depolarization effect,
since the predominantly polarized emission regions are compact enough
to be covered by the resolutions.
Therefore the best fit result seems to indicate that
(a) any existing Faraday screen does not significantly affect
the polarized emission of 3C~279 at mm and sub-mm,
or (b) a Faraday screen contains a uniform field,
even if there is a Faraday rotation by the Faraday screen~\citep{far+14}.
In fact, the estimated $|{\rm RM}|$ of 647-2713 rad m$^{-2}$ over the wavelengths
3.5-13 mm may indicate that the polarized emission at the wavelengths
is passing through a Faraday screen which may contain a uniform field.
However,
due to the resolution limitation,
the observations only probe the overall direction of the magnetic field
in the mm emission-dominated region.
Obviously, 
in the future, short-millimeter (and sub-millimeter)
VLBI observations~\citep{fis+13,til+14}
are much awaited to
really probe the geometry of the magnetic field in detail.

\cite{jor+07} reported that RM at shorter wavelengths is larger in relativistic
radio jet with a wavelength dependence of RM
: $|{\rm RM}(\lambda)|\propto\lambda^{-a}$,
assuming that the Faraday rotation originates in or in close proximity of the jet.
For optically thick VLBI cores, where the optical depth is unity,  
a simple jet model is assumed as the following~\citep[see also][]{lob98}:
the distance $r$ of the emission region to the central engine depends on 
the observing wavelength $\lambda$ as $r\propto\lambda$,
the electron density $n_{\rm e}$
and the magnetic field parallel to the line of sight $B_{||}$
in the region scale geometrically as $n_{\rm e}\propto r^{-2}$
(for the spherical or conical geometries of the jet)
and $B_{||}\propto r^{-1}$, and the path length $l$ increases
as a function of $r$ as $l\propto r$.
The RM dependence on the electron density, the parallel magnetic field strength,
and the path length ${\rm RM}\propto\int n_{\rm e}B_{||}dl$ gives
$|{\rm RM}(\lambda)|\propto\lambda^{-2}$.
The authors found $a={1.8\pm0.5}$ for eight AGNs
observed at centimeter to mm wavelengths. 
This wavelength dependence of RM was confirmed for 1418+546
with $a=1.9\pm0.3$~\citep{tri+12a},
for 3C~84 with $a=2$~\citep{far+14},
and for PKS~1830-211 with $a\approx2.4$~\citep{mart+15},
although a larger value of $a=3.6\pm1.3$ was reported~\citep{alg13}.
For 3C~279, we found that the 7-13~mm $|{\rm RM}|$ of $6.5\times10^2$ rad m$^{-2}$
can be scaled
by $\lambda^{-2.2}$ (i.e., $a=2.2$)
to get $|{\rm RM}|\sim2.7\times10^3$ rad m$^{-2}$ at 3.5-7 mm,
which is consistent with the measured value.
This may imply that the Faraday rotation at mm wavelength originate in
or in the close proximity of the jet whose geometry is spherical or conical.
Caution should be taken here,
as this model may not fully apply for the mm emission from 3C 279,
since the presence of knots (e.g., shocks) such as
C1 and C2 components, visible at mm wavelengths,
may indicate that the innermost jet is composed of both
smooth and knotty components~\citep[e.g.,][]{kud+11}.

The first detection of $350~\mu$m polarized emission of 3C~279,
the multiwavelength mm polarization measurements, and 
the high resolution multiwavelength VLBI observations on mas scales
enable us to find that the mm and sub-mm polarization emission
of 3C~279 are generated in the compact jet regions within 1 mas scale
and affected by a Faraday screen in or in the close proximity
of the jet whose geometry is spherical or conical.
We also found that the dominant magnetic field direction in the region
is perpendicular to the direction of the mm jet at mas scale.

\acknowledgments
We thank the
CSO staff for obtaing the $350\mu$m data,
and the KVN staff
for operating the array and correlating the data.
The KVN is a facility operated by
the Korea Astronomy and Space Science Institute.
The KVN operations are supported
by Korea Research Environment Open NETwork
which is managed and operated
by Korea Institute of Science and Technology Information.
We thank Mark Gurwell for providing the data obtained at
the SMA, which
is a joint project between the Smithsonian
Astrophysical Observatory and the Academia Sinica Institute of Astronomy
and Astrophysics and is funded by the Smithsonian Institution and
the Academia Sinica.
Data from the Steward Observatory spectropolarimetric monitoring project
which is supported by Fermi Guest Investigator
grants NNX08AW56G, NNX09AU10G, and NNX12AO93G, were used.


\end{document}